\documentclass{ifacconf}

\usepackage{graphicx}      
\usepackage{natbib}
\usepackage{amsmath}
\usepackage{amssymb}
\begin{document}
\begin{frontmatter}

\title{Estimating and detecting random processes on the unit circle} 

\author[First]{Changrong Liu} 
\author[Second]{S. Suvorova} 
\author[Second]{R. J. Evans}
\author[Second]{W. Moran}
\author[Third]{A. Melatos}

\address[First]{Electrical and Electronic Engineering Department, University of Melbourne, Parkville, Victoria 3010, Australia and OzGrav (e-mail: changrongl1@student.unimelb.edu.au)}
\address[Second]{Electrical and Electronic Engineering Department, University of Melbourne, Parkville, Victoria 3010, Australia and OzGrav (e-mail: sofia.suvorova, robinje, wmoran@unimelb.edu.au).}
 \address[Third]{School of Physics, University of Melbourne, Parkville, Victoria 3010, Australia and OzGrav (e-mail: amelatos@unimelb.edu.au).}

\begin{abstract}                
Detecting random processes on a circle has been studied for many decades. The Neyman-Pearson detector, which evaluates the likelihood ratio, requires first the conditional mean estimate of the circle-valued signal given noisy measurements, which is then correlated with the measurements for detection. This is the estimator-correlator detector. However, generating the conditional mean estimate of the signal is very rarely solvable. In this paper, we propose an approximate estimator-correlator detector by estimating the truncated moments of the signal, with estimated signal substituted into the likelihood ratio. Instead of estimating the random phase, we estimate the complex circle-valued signal directly. The effectiveness of the proposed method in terms of estimation and detection is shown through numerical experiments, where the tracking accuracy and receiver operating curves, compared with the extended Kalman filter are shown under various process/measurement noise.
\end{abstract}


\end{frontmatter}

\section{Introduction}
Detecting weak circle-valued random processes in white Gaussian noise has been analyzed by researchers for several decades. This type of signal is widely encountered in a range of fields including communication systems, where signals are modelled as either frequency or phase modulated by a Gaussian message (\cite{1054093}). Another example is optical communication with laser phase drift (\cite{rusch1995effects}; \cite{foschini1988noncoherent}). Generally speaking, These signals are specific cases of non-Gaussian random processes (\cite{kailath1998detection}). In order to construct a Neyman-Pearson detector for such a process, the likelihood ratio must be evaluated, which requires computation of the causal conditional mean  estimate of the random process given measurements. The estimated signal is then correlated  with the measurements to produce the detection statistic (\cite{kailath1998detection}; \cite{veeravalli1991quadratic}). This is the estimator-correlator (EC) detector (\cite{kailath1998detection}). Unfortunately, the conditional mean estimate is mostly impossible to compute except in the Gaussian case (\cite{kailath1998detection}). Instead, a near-optimal approximation of the conditional mean estimate (or equivalently, the conditional density) of the signal has to be constructed.

For circle-valued signals, various approximation techniques have been proposed. A random point on the circle is represented either by the angle $\Theta\in [-\pi,\pi)$ or the complex numbers $\{e^{i\Theta}|\Theta\in [-\pi,\pi)\}$. One approach is to characterize the conditional density of $\Theta$ using a Gaussian sum approximation with the mean and variance updated, as described in \cite{1093926}. In \cite{1055280}, the density of $\Theta$ is spanned by Fourier modes, with Fourier coefficients estimated.  In \cite{9382922}, the density of $\Theta$ is approximated by a von-Mises distribution with maximum entropy, and at each iteration, the density is propagated and projected back to a von-Mises distribution. Another approach is to use the extended Kalman filter (EKF) to linearize the nonlinear system dynamics and/or the measurement process to demodulate (estimate) the phase (\cite{DBLP:journals/tcom/GeorghiadesS85}; \cite{1054093}). In \cite{1057115}, the unwrapped phase model is assumed and fitted by a linear regression. In \cite{1056241}, the dynamics of the wrapped phase sequence is assumed and then estimated using dynamic programming. A similar idea has also been proposed in \cite{8501578} and \cite{PhysRevD.104.042003}, where the phase is modelled as a hidden Markov model and the Viterbi algorithm is used to track the random phase. A deterministic filtering approach is attempted in \cite{article}, where the process and measurement noises are both treated as unknown deterministic disturbances and a deterministic minimum-energy filter is formed by minimizing an energy cost function over phase trajectories (\cite{5399999}). As we can see, a large number of methods have been considered for estimating (filtering) the phase $\Theta$, while very few papers analyze the complex signal $e^{i\Theta}$ directly. One recent paper (\cite{https://doi.org/10.48550/arxiv.2108.02602}) minimizes a Tiknonov-regularized problem over the complex signal $e^{i\Theta}$, where the non-convexity introduced by the $S^1$ constraint is relaxed using moments of measures. 

In this paper, we focus directly on stochastic filtering of the circle-valued signal $e^{i\Theta}$ embedded in large additive white noise, where $\Theta$ obeys different random processes.  We first present the nonlinear Kushner stochastic partial differential equation for recursively computing the conditional mean estimate of the signal, or equivalently, the conditional probability density.  The solution of this equation is infinite dimensional, and we approximate it by a sequence of moment functions. The choice of moment functions is determined by the fact that they form the eigenspace of the infinitesimal generator of the signal process. In doing so, solving the Kushner equation simplifies to solving a sequence of time-dependent ordinary differential equations (ODEs), which characterize the evolution of the conditional density. We then produce the approximate optimal EC detector and show its performance by simulations.

The paper is organized as follows. Section \ref{section II} describes the dynamic system on $S^1$ and formulates the detection and estimation problems. The Kushner equation for computing the conditional density  and its finite-dimensional approximation is introduced and derived in Section \ref{section III}. Simulation results are included in Section \ref{section IV}, testing both estimation and detection performance, compared with EKF based methods. Finally, brief conclusions are drawn in Section \ref{section V}.
\section{Problem Formulation}
\label{section II}
\subsection{System model}
Consider a detection system with two hypothesis:
\begin{align}
    H_0: Z_t&=\sigma_0^{1/2}W_t, \;0\leq t \leq T, \label{H0}\\
    H_1: Z_t&=\int_0^t h(X_t) ds+\sigma_0^{1/2}W_t, \; 0\leq t \leq T \label{H1}
\end{align}
where $Z_t$ is the noisy measurements and $X_t=e^{i\Theta_t}\in S^1$ is the circle-valued signal with $\Theta_t$ obeying an arbitrary random process. Measurement function $h(X_t)$ in our case, is given by 
\begin{equation}
\label{measurement function}
    h(X_t)={\Re}[X_t]
\end{equation}
with $\Re[\cdot]$ denoting the real part. $\{W_t,0\leq t \leq T\}$ is the standard Wiener process, which represents the measurement noise and is independent of $X_t$. $\sigma_0$ denotes the variance of the measurement noise and is assumed known. In this paper, we consider three specific random processes for $\Theta_t$:
\begin{align}
    \text{(i)} \; \Theta_t &= \int_0^t q_{\Theta}^{1/2}dB_s+\phi_0, \label{phase mod1}\\
    \text{(ii)} \; \Theta_t&=\int_0^t (w_0ds+q_{\Theta}^{1/2}dB_s) +\phi_0 \label{phase mod2},\text{ and}\\
    \begin{split}
    \text{(iii)} \; \Theta_t&=\int_0^t (w_sds +q_{\Theta}^{1/2}dB_s)+\phi_0,\\
    w_t&=w_0+\int_0^t q_w^{1/2}d\tilde{B}_{s}\label{phase mod3}
    \end{split}
\end{align}
where $q_{\Theta}$ and $q_w$ are known functions of  phase, $\Theta$,  and frequency, $w$. They represent the noise variance for phase and frequency, respectively. $w_0$ is the known initial frequency and $\phi_0$ is the random initial phase, uniformly distributed across $[0,2\pi)$. $\{B_t,0\leq t \leq T\}$ and $\{\tilde{B}_t,0\leq t \leq T\}$ are two independent standard Wiener processes, which are also independent of $X_t$ and $\{W_t,0\leq t \leq T\}$.
\subsection{Neyman-Pearson detector}
The Neyman-Pearson (EC) detector is the optimal detector in the sense that it maximizes the detection probability (Pd) with given false alarm probability (Pf). In order to form this type of detector, the log likelihood ratio $\log \Lambda_t$ has to be evaluated, which is given by
\begin{align}
\log \Lambda_t &= \int_0^t \hat{X}_sdZ_s-\frac{1}{2}\int_0^t \hat{X}_s^2ds, \label{log likelihood}\\
\hat{X}_t&=\mathbf{E}[X_t|\mathcal{F}_t;H_1]\label{xhat}
\end{align}
with $\mathcal F_t$ the filtration of $\sigma$-algebras generated by $\{Z_\tau,0\leq \tau \leq t\}$, given in (\ref{H1}) with $\mathcal{F}_s\subseteq \mathcal{F}_t$, $\forall s \leq t$.

From (\ref{log likelihood}) and (\ref{xhat}), the EC detector is formed in two steps: 1.) calculate the conditional mean estimate $\hat{X}_t$; 2.) form the log likelihood ratio $\log \Lambda_T$ at the termination time $T$ and compare it with a pre-specified threshold to claim $H_0$ or $H_1$ correspondingly. 
\section{Nonlinear Filtering: compute conditional mean estimate}
\label{section III}
We now consider computation of the conditional mean estimate of the signal using the Kushner equation.
\subsection{Kushner equation: recursive filtering of the conditional density}
Given the filtered probability space $(\Omega,\mathcal{F},\mathbb{P};\{\mathcal{F}_t\})$ with $\Omega$ the underlying sample space and $\mathbb{P}$ the probability measure defined on $\mathcal{F}$, suppose the signal and measurement processes are given by
\begin{equation}
\label{general system}
\begin{split}
    dx_t&=b(x_t)dt+\sigma(x_t)dB_t ,\: x(0)=x_0\\
    dZ_t&=h(x_t)dt+\sigma_0^{1/2} dW_t,\: Z_0 =0 
\end{split}
\end{equation}
with $x_t\in \mathbb{R}^d$ and $b: \mathbb{R}^d\to  \mathbb{R}^d$, $\sigma:  \mathbb{R}^d\to  \mathbb{R}^{d\times p}$, $h:  \mathbb{R}^d\to  \mathbb{R}$, $a\overset{\triangle}{=}\frac{1}{2}\sigma\sigma^{\rm{T}}$ with superscript $\rm{T}$ being the transpose of a matrix.  The filtering problem can be formulated as: for a general measurable function $g(x_t)$, calculate $\hat{g}(x_t)$ by
\begin{equation}
\label{ghat}
\begin{split}
    \hat{g}(x_t)&\overset{\triangle}{=}\mathbf{E}[g(x_t)|\mathcal{F}_t]\\
    &=\int_{\mathbb{R}^d} g(x_t)\mathbb{P}(x_t\in dx|\mathcal{F}_t)\\
    &=\int_{\mathbb{R}^d} g(x)p(x,t|\mathcal{F}_t)dx
\end{split}   
\end{equation}
where $p(x,t|\mathcal{F}_t)$ denotes the conditional density of $x$ at time $t$. The recursive formula for evolving  $p(x,t|\mathcal{F}_t)$ is described by the  Kushner equation (\cite{kushner1967nonlinear})
\begin{equation}
\label{dp}
\begin{split}
    dp(x,t|\mathcal{F}_t)&=\mathcal{L}^*p(x,t|\mathcal{F}_t)dt \\
    &+(h(x_t)-\hat{h}(x_t))\cdot \sigma_0^{-1}[dZ_t-\hat{h}(x_t)dt]p(x,t|\mathcal{F}_t)
\end{split}
\end{equation}
with $\mathcal{L}^*$ the adjoint of the infinitesimal generator $\mathcal{L}$, given respectively by 
\begin{align}
    \mathcal{L}^* (\cdot)&=\sum_{i,j=1}^d \frac{\partial^2}{\partial x_i \partial x_j}(a_{ij}(\cdot))-\sum_{i=1}^d \frac{\partial}{\partial x_i}(b_i (\cdot)),\\
        \mathcal{L}(\cdot)&=\sum_{i,j=1}^d a_{ij}\frac{\partial^2 }{\partial x_i x_j}+\sum_{i=1}^d b_i\frac{\partial }{\partial x_i}.\label{L}
        \end{align}
To uniquely characterize the conditional density, we pair (\ref{dp}) with an arbitrary test function $g(x)$ with compact support, where the duality pairing is denoted by $\langle\cdot,\cdot\rangle$. We then have
\begin{align}
    d\Big\langle p(x,t|\mathcal{F}_t),g(x)\Big\rangle=&\Big\langle\mathcal{L}^*p(t,x|\mathcal{F}_t),g(x)\Big\rangle dt +\Big\langle(h-\hat{h})\nonumber\\
    &\cdot \sigma_0^{-1}[dZ_t-\hat{h}dt]
    p(x,t|\mathcal{F}_t),g(x)\Big\rangle
\end{align}
which further implies 
\begin{equation}
\begin{split}
\label{ghat2}
    d\hat{g}(x_t) =& \mathcal{L}\hat{g} (x_t) dt\\
    &+ \langle g(x_t)h(x_t)\rangle-\hat{g}(x_t)\hat{h}(x_t)\cdot \sigma_0^{-1}[dZ_t-\hat{h}(x_t)dt],
    \end{split}
\end{equation}
here $\mathcal{L}$ is defined in (\ref{L}). Both $\hat{}$ and $\langle \cdot \rangle$ denote the conditional mean with respect to the conditional density $p(x,t|\mathcal{F}_t)$. $\hat{g}(x_t)\overset{\triangle}{=}\Big\langle p(x,t|\mathcal{F}_t),g(x)\Big \rangle$ is defined in (\ref{ghat}). 


 \subsection{Choice of the test function}
 From the previous section,  we observe the duality relationship between the conditional density and the estimated test function. In order to fully characterize $p(x,t|\mathcal{F}_t)$, an infinite number of statistics (or test functions $g(\cdot)$) are required. In this paper, we set the test functions to be the moment functions of the signal $X_t$, given as $X_n(t)\overset{\triangle}{=}e^{in\Theta_t}$ for $n=0\cdots, N-1$, considering firstly, $X_n(t)$ can be regarded as the characteristic function of $\Theta$, which completely determines the properties of the probability distribution of $\Theta$. Secondly, $X_n(t)$ spans the eigenspace of the generator $\mathcal{L}$, as discussed in the following section. In doing so, solving an infinite dimensional Kushner equation boils down to solving $N$'s ODEs.
 
 As we can see from (\ref{ghat2}), the filtering process is composed of two terms: the one-step prediction (the first term) and the update (the second term). We will derive the solutions for $\hat{X}_n$ in terms of these two properties.
\subsection{Prediction: compute $\mathcal{L}\hat{X}_n(t)$}
In order to obtain the infinitesimal generator of the signal process, we need to write the signal dynamics in Ito differential form, for three scenarios. In particular, we have for scenario (i) and (ii)
\begin{align}
   {\text{(i)}} \; dX_n&=-\frac{q_{\Theta}n^2}{2}X_ndt+inq_{\Theta}^{1/2}X_n dB_t,
    \label{dynamics1}\\
    {\text{(ii)}}\; dX_n&=(inw_0-\frac{q_{\Theta}n^2}{2})X_ndt+inq_{\Theta}^{1/2}X_n dB_t.
    \label{dynamics2}
\end{align}
Comparing (\ref{dynamics2}) with (\ref{dynamics1}), a constant rotation term $inw_0$ is added to the drift coefficient. Considering the martingale process in both circumstances: $inq_{\Theta}^{1/2}X_n dB_t$, with the property $\mathbf{E}[\int_s^t inq_{\Theta}^{1/2}X_n(\tau) dB_\tau|\mathcal{F}_s]=0$, we have a separate infinitesimal generator for each $X_n$, where $\mathcal{L}_n = inw_0-\frac{q_{\Theta}n^2}{2}$, with $w_0=0$ in scenario (i).

The differential equation for $X_n$ in scenario (iii) can be obtained in the same manner
\begin{equation}
        dX_n=(inw_t-\frac{q_{\Theta}n^2}{2})X_ndt+inq_{\Theta}^{1/2}X_n dB_t.
    \label{FM}
\end{equation}
However, as shown here, the randomness appears in the drift coefficient. Now it is difficult to interpret the generator by simply writing $\mathcal{L}_n(w) = inw(t)-\frac{q_{\Theta}n^2}{2}$ with $w(t)$ a random process. Instead, $\mathcal{L}_n(w)$ should be interpreted in the mean sense by marginalizing out the random variable $w(t)$. In other words, we have the infinitesimal generator computed by $\mathcal{L}_n =\mathbf{E}_w[\mathcal{L}_n(w)|\mathcal{F}_t]= \int_{w\in \mathbb{R}} \mathcal{L}_n(w)p(w,t|\mathcal{F}_t)dw$, with $d\hat{X}_n-\mathcal{L}_n \hat{X}_ndt$ still a martingale process. Alternatively, we introduce the new state variables as $X_{mn}\overset{\triangle}{=}w^m e^{ in\Theta}$, with $m=0,\dots,M-1$ and $n=0,\dots N-1$, where superscript $m$ and $n$ denote the $m$th and $n$th power, respectively. This is an ``augmented" state space by noting that $X_n = X_{mn}|_{m=0}$. For predicting $X_{mn}$, likewise, we write down the differential equation for $X_{mn}$ as
\begin{equation}
\begin{split}
    {\text{(iii)}}\;dX_{mn}
    =&\Big(\frac{q_wm(m-1)}{2}X_{m-2,n}+inX_{m+1,n}\\
    &-\frac{q_{\Theta}n^2}{2}X_{mn}\Big)dt
    +q_w^{1/2}mX_{m-1,n}d\tilde{B}_t\\
    &+inq_{\Theta}^{1/2}X_{mn}dB_t
    \label{dynamics3}
    \end{split}
\end{equation}
with $\mathcal{L}_{mn}(X_{mn})=\frac{q_wm(m-1)}{2}X_{m-2,n}+inX_{m+1,n}-\frac{q_{\Theta}n^2}{2}X_{mn}$ for each $X_{mn}$. Here by augmenting the state space from $X_n$ to $X_{mn}$,   the time-variant generator is replaced by a time-invariant generator at the cost of higher computational complexity.
\subsection{Update: incorporate the innovation $dZ_t-\hat{h}dt$ }
In all three scenarios, we have measurement function $h(X_t)=\Re[X_t]$ from (\ref{measurement function}). This is a considerable advantage since $h$ can be regarded as a linear operator in both the subspace $\{X_n\}_{n=0}^{N-1}$ and $\{X_{mn}\}_{m=0,n=1}^{M-1,N-1}$ by noting
\begin{align}
    \begin{split}{\text{(i)}}\& \text{(ii)}  \;  \langle hX_n\rangle&=\Big\langle \frac{X_{n+1}+X_{n-1}}{2}\Big\rangle\\
    \hat{h}&=\frac{\hat{X}_1+\hat{X}_{1}^*}{2}, \label{hhat}\end{split}\\
    \begin{split}{\text{(iii)}}\;     \langle hX_{mn}\rangle&=\Big \langle \frac{X_{m,n+1}+X_{m,n-1}}{2}\Big \rangle\\
    \hat{h}&=\frac{\hat{X}_{01}+\hat{X}_{01}^*}{2}
   \label{hhat2} \end{split}
\end{align}
with $()^*$ denoting the complex conjugation. Combining (\ref{dynamics1}), (\ref{dynamics2}), (\ref{dynamics3}) with (\ref{hhat}), (\ref{hhat2}) and substituting into (\ref{ghat2}), we can compute $\hat{X}_n(t)$ for (i)\&(ii) and $\hat{X}_{mn}(t)$ for (iii) recursively by solving a sequence of stochastic ODEs.

The log-likelihood ratio, defined in (\ref{log likelihood}) can then be updated recursively as well by substituting in $\hat{X}_t$ from (\ref{ghat2}) without too much effort, considering $\hat{X}_t=\hat{X}_n(t)|_{ n=1}$ for (i)\&(ii) and $\hat{X}_t=\hat{X}_{mn}(t)|_{ m=0,\;n=1}$ for (iii).
\section{Simulations and results}
\label{section IV}
\subsection{Implementation of the Neyman-Pearson (EC) detector}
Since scenario (i) is analogous to (ii) with $w_0=0$, we only do experiments for (ii) and (iii). The Neyman-Pearson (EC) detector is composed of a conditional mean estimator and a log-likelihood ratio detector, hence we need to check the performance of both estimation and detection. For estimation, we generate the synthetic signal $X_{\rm syn}=A_{\rm syn}e^{i\Theta_{\rm syn}}$ with $\Theta_{\rm syn}$ given in (\ref{phase mod2}) and (\ref{phase mod3}). Noisy measurements are generated from (\ref{H1}).  For detection, we randomly generate measurements from (\ref{H0}) or (\ref{H1}), where signal $X_{\rm syn}=A_{\rm syn}e^{i\Theta_{\rm syn}}$ is generated as above. We fix termination time $T$ and compare $\log \Lambda_T$ with a pre-specified threshold and claim detection if $\log \Lambda_T$ is greater than the threshold, and vice versa. We simulate both the signal process and the measurement process numerically using the Euler-Maruyama method with sampling interval $\Delta t =0.1$s. The parameters are listed in Table \ref{tab: parameter description }.

\begin{table}[hb]
\centering
\caption{Simulation parameters}
\label{tab: parameter description }
\begin{tabular}{ |p{2.5cm}|p{3cm}|p{2.1cm}|} 
\hline
 parameter &  description& value    \\ 
 \hline
 $\phi_0$ & initial phase & $\overset{\rm dist}{\sim} U(0,2\pi)$   \\ 
  \hline
 $q_{\Theta}$& phase noise variance&$10^{-1},10^{-2}$\\
  \hline
 $q_{w}$& frequency noise variance&$10^{-8}$\\
  \hline
 $\sigma_0$ & observation noise variance &$1,\dots,80$\\
  \hline
 $w_0$ & initial frequency & $0.012,\;0.032$Hz\\
  \hline
 $|A_{\rm{syn}}|$ & absolute amplitude of synthetic signal &$1$\\
  \hline
SNR $=\sqrt{\frac{|A_{\rm{syn}}|\Delta t}{\sigma_0}}$  & signal to noise ratio & $0.0354,\ldots 0.3162$\\
 \hline
 $N-1$& highest moment of $X_t$&$11$\\
  \hline
 $M-1$& highest moment of $w_t$&$3$\\
  \hline
 $\Delta t$&sampling time interval&$10^{-1}$ sec\\
  \hline
 $L$& length of synthetic data& $ 10^{4}$\\
  \hline
 $T=L\Delta t$ & termination time & $ 10^3$ sec\\
 \hline
\end{tabular}
\end{table}
\subsection{Estimation performance}
\begin{figure}[!ht]
\begin{center}
\includegraphics[width=8.4cm]{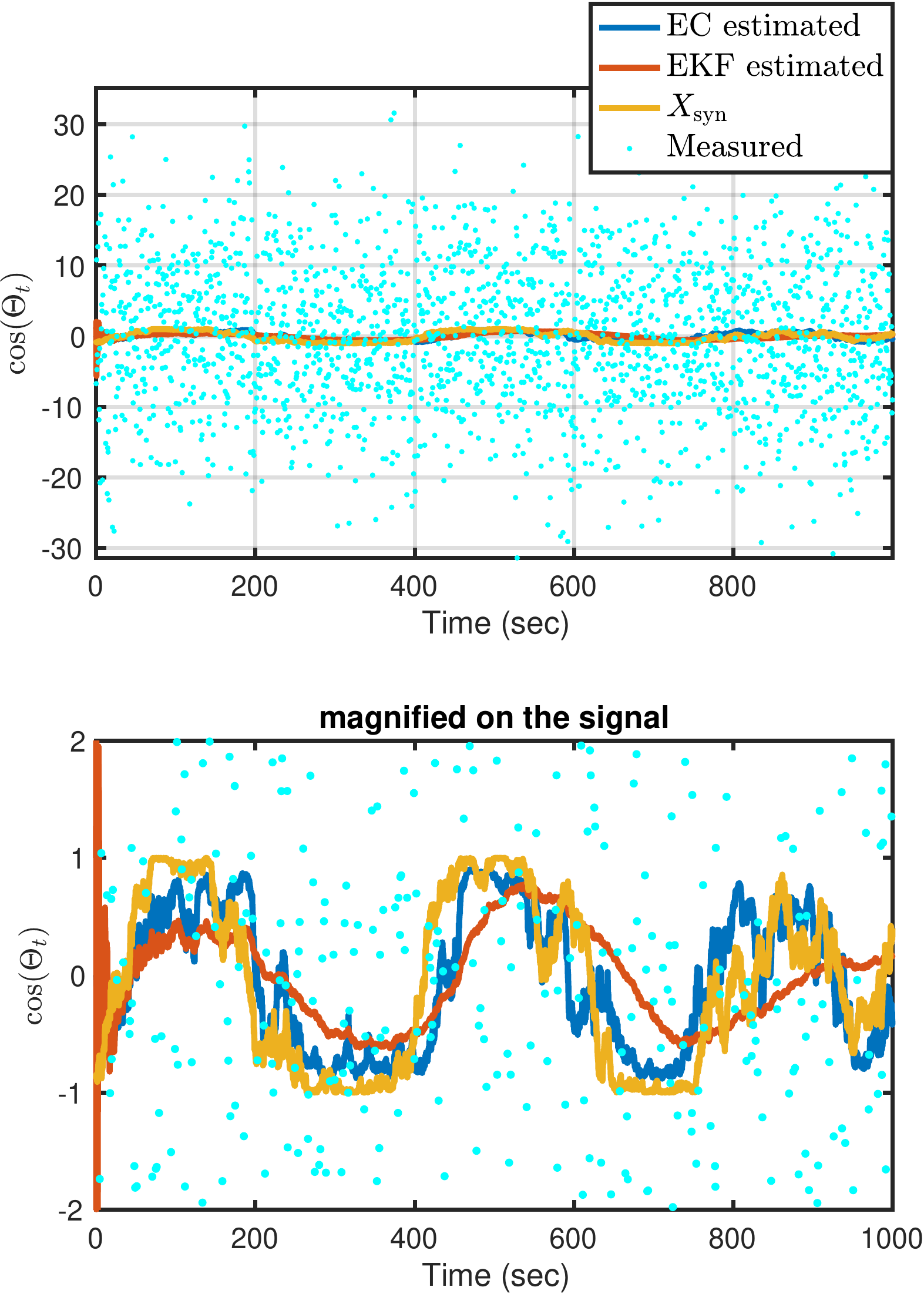}    
\caption{Conditional mean estimate under scenario (ii): Plot of $\Re[\hat{X}_t]$ estimated by EC (blue) and EKF (red) at SNR = 0.1 with $\Theta_t$ defined in (\ref{phase mod2}), where $w_0=0.012$Hz and $q_{\Theta}=10^{-2}$. The bottom panel is a magnified version of the top.} 
\label{fig:xhat1}
\end{center}
\end{figure}

\begin{figure}[!ht]
\begin{center}
\includegraphics[width=8.4cm]{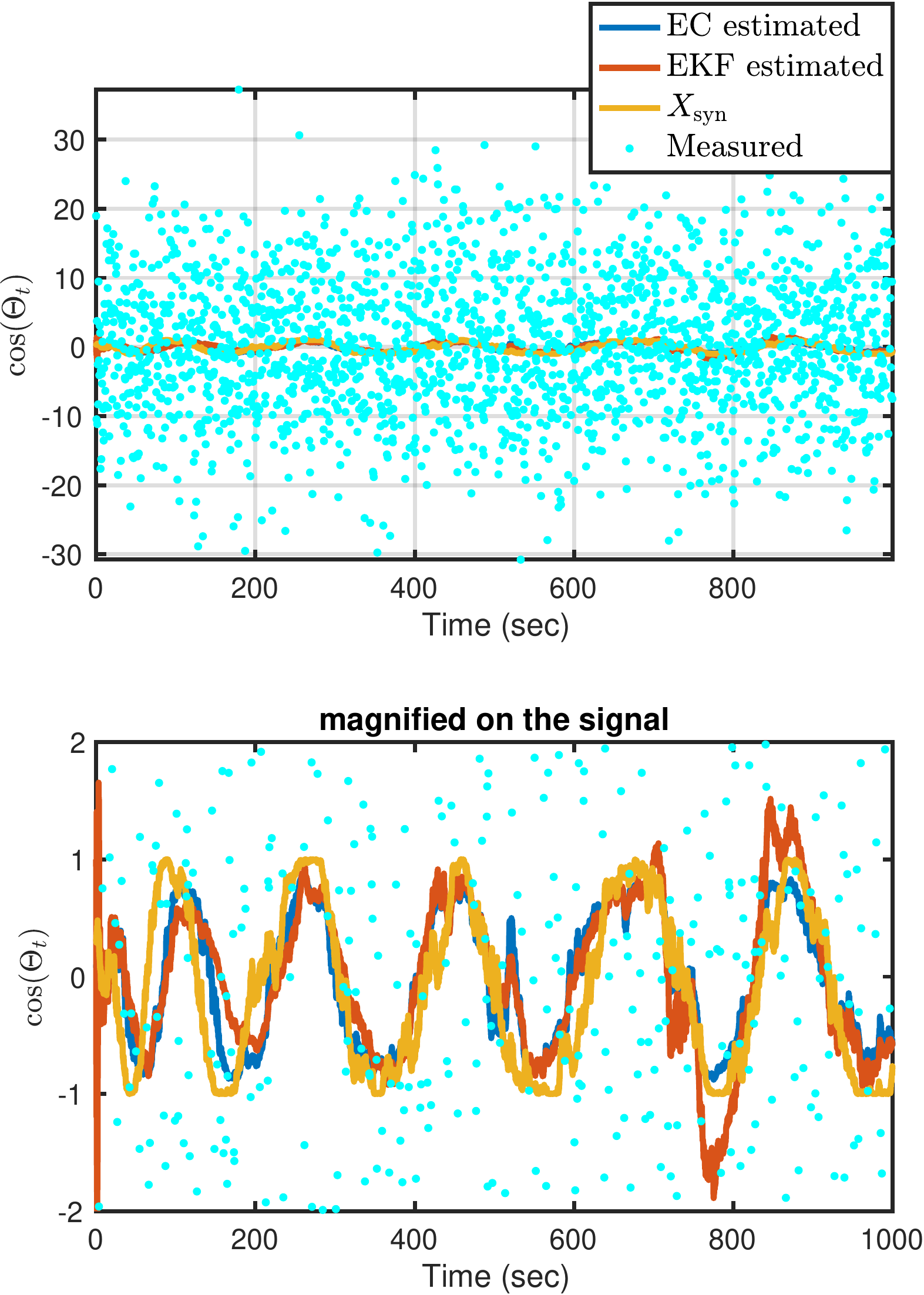}    
\caption{Conditional mean estimate under scenario (iii): Plot of $\Re[\hat{X}_t]$ estimated by EC (blue) and EKF (red) at SNR = 0.1 with $\Theta_t$ defined in (\ref{phase mod3}), where $w_0=0.032$Hz, $q_{\Theta}=10^{-2}$ and $q_w=10^{-8}$. The bottom panel is a magnified version of the top.} 
\label{fig:xhat2}
\end{center}
\end{figure}

 Two realizations of EC estimated $\Re[\hat{X}_t]$, with phase dynamics described in (ii) and (iii) are shown in Fig.~\ref{fig:xhat1} and Fig.~\ref{fig:xhat2}, compared with EKF estimated $\Re[\hat{X}_t]$. From both plots, with SNR = 0.1, the signal is submerged in large noise. However, the EC estimated signal (blue) is closer to the synthetic signal (yellow) compared with EKF estimated signal (red). When frequency experiences small wandering in scenario (iii), as displayed in Fig.~\ref{fig:xhat2}, by estimating the product $X_{mn}$, we can still extract $\hat{X}_t$ with higher accuracy than EKF estimated signal.

\subsection{Detection performance}
\begin{figure}[!ht]
\begin{center}
\includegraphics[width=8.4cm]{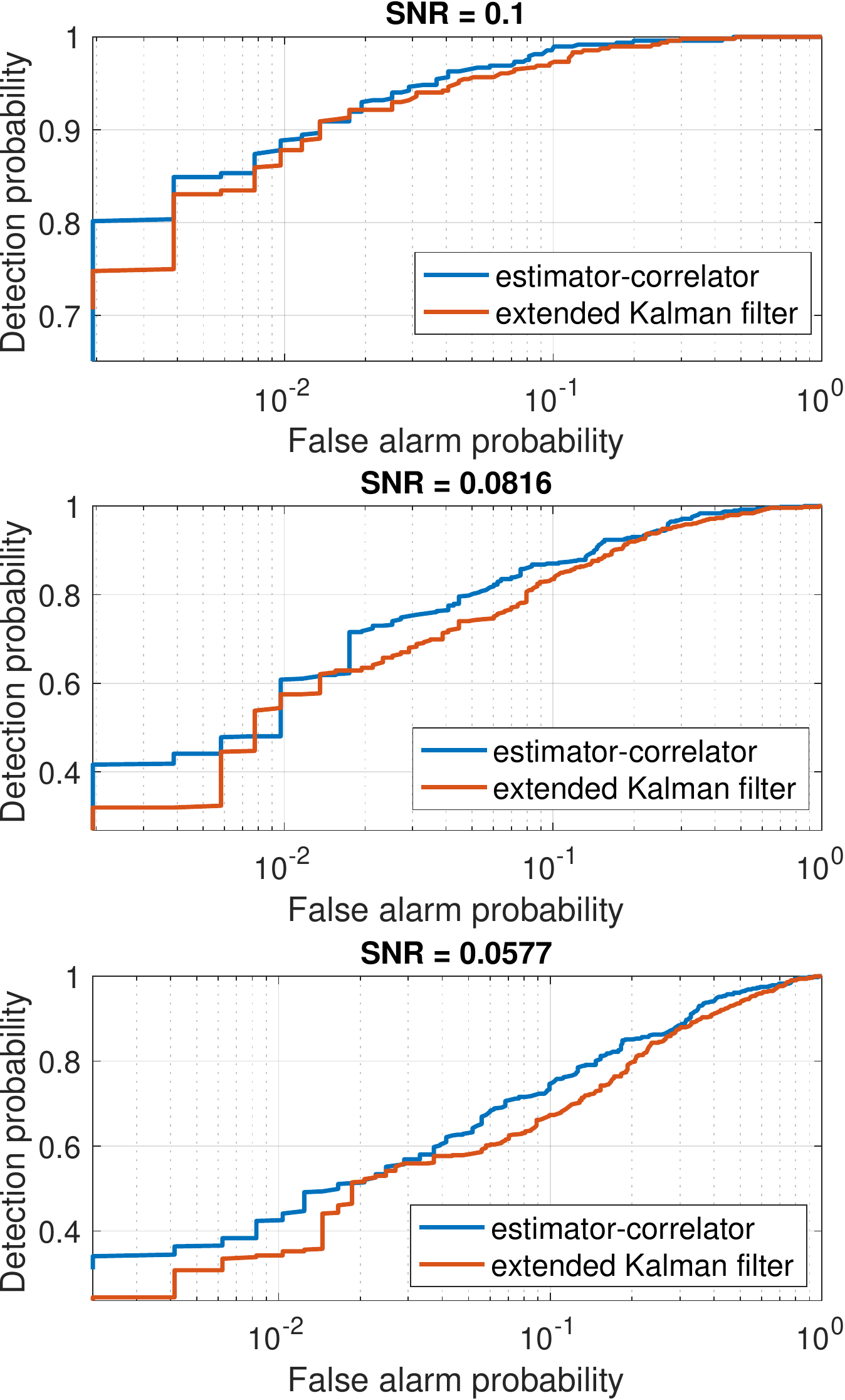}    
\caption{ROC curves under scenario (ii): false alarm probability vs detection probability as SNR drops from 0.1 (top) to 0.0816 (middle) then to 0.0577 (bottom), with $q_\Theta=10^{-1}$ and $w_0=0.012$Hz.} 
\label{fig:roc1}
\end{center}
\end{figure}

\begin{figure}[!ht]
\begin{center}
\includegraphics[width=8.4cm]{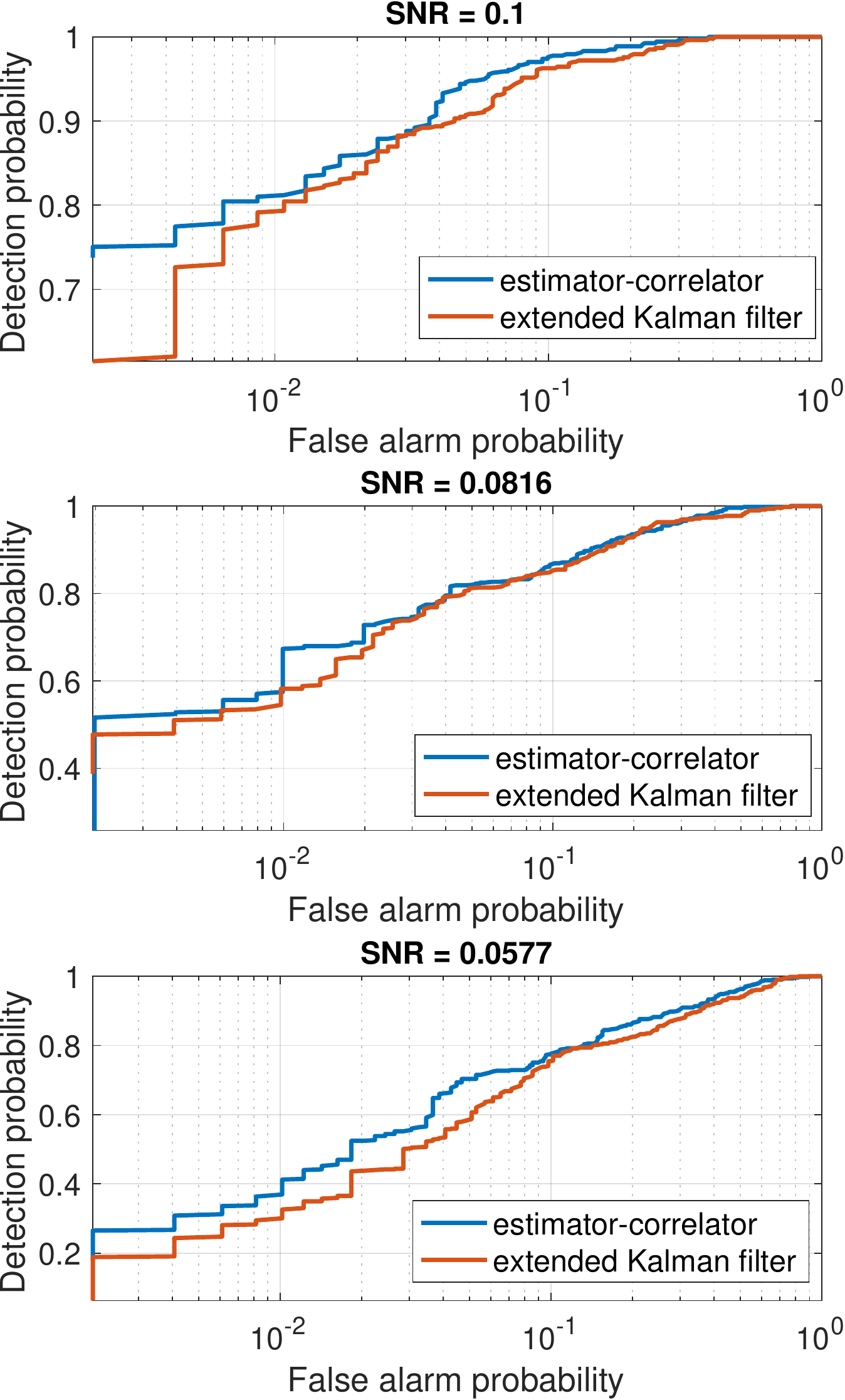}   
\caption{ROC curves under scenario (iii): false alarm probability vs detection probability as SNR drops from 0.1 (top) to 0.0816 (middle) then to 0.0577 (bottom), with $q_\Theta=10^{-1}$, $q_w=10^{-8}$ and $w_0=0.012$Hz.} 
\label{fig:roc2}
\end{center}
\end{figure}

To quantify the performance of the approximate EC detector, receiver operating characteristic (ROC) curves with different SNRs under scenario (ii) with $q_{\Theta}=10^{-1}$ and $w_0=0.012$Hz are displayed in Fig. \ref{fig:roc1}, compared with the EKF detector as well. In Fig.~\ref{fig:roc2}, ROCs of the EC and EKF detector under scenario (iii) with $q_{\Theta}=10^{-1}$, $q_w=10^{-8}$ and $w_0=0.012$Hz are plotted.  From Fig.~\ref{fig:roc1}, as SNR drops from 0.1 (top) to 0.0816 (middle) and then to 0.0577 (bottom), Pd of the EC detector (blue) descends from 0.9 to 0.6, then to 0.4 at  Pf $=10^{-2}$.  Fig.~\ref{fig:roc2} exhibits a similar trend, with higher Pd as SNR gets larger. Comparing two top panels between Fig.~\ref{fig:roc1} and Fig.~\ref{fig:roc2}, when the frequency is slowly wandering, with $q_w=10^{-8}$ in Fig.~\ref{fig:roc2}, EC experiences a slight degradation, with Pd drops from 0.9 to 0.8 at SNR = 0.1 and Pf = $10^{-2}$. Throughout all six panels, the EC detector advantages over the EKF detector with higher Pd across the Pf range, especially when Pf $<10^{-2}$.

\section{Conclusion}
\label{section V}
In this paper, we focus on estimating and detecting random processes on the circle. The structure of the Neyman-Pearson (or EC) detector is split into two parts: a conditional mean estimate and a log-likelihood ratio detector. The conditional mean estimate, which is described by the nonlinear Kushner stochastic partial differential equation is approximated by filtering the first $N$ moments of the circle-valued signal. Instead of approximating the conditional density of the phase, we characterize the conditional density of the circle-valued signal directly, by pairing it with the moments of the signal. When the frequency is also a random process, in addition to filtering the moment function of the signal, we estimate the product of moment functions for frequency and signal, resulting in a deterministic generator.  

We perform Monte Carlo simulations to verify the estimation performance by displaying realizations of the EC and EKF estimated signals and  plotting the ROC curves for the detection, comparing also with the EKF detector. Overall, EC estimated signal has better tracking (estimating) capacity and higher detection probability than EKF based methods.

\begin{ack}
The authors acknowledge support from the Australian Research Council (ARC) through the Centre of Excellence for Gravitational Wave Discovery (OzGrav) (grant number CE170100004) and an ARC Discovery Project (grant number DP170103625).
\end{ack}

\bibliography{ifacconf}             
                                                   







\end{document}